\documentclass[twoside,twocolumn,9pt]{article}
\usepackage{extsizes}
\usepackage[super,sort&compress,comma]{natbib} 
\usepackage[version=3]{mhchem}
\usepackage[left=1.5cm, right=1.5cm, top=1.785cm, bottom=2.0cm]{geometry}
\usepackage{balance}
\usepackage{mathptmx}
\usepackage{sectsty}
\usepackage{graphicx} 
\usepackage{lastpage}
\usepackage[format=plain,justification=justified,singlelinecheck=false,font={stretch=1.125,small,sf},labelfont=bf,labelsep=space]{caption}
\usepackage{float}
\usepackage{fancyhdr}
\usepackage{fnpos}
\usepackage[english]{babel}
\addto{\captionsenglish}{%
  
}
\usepackage{array}
\usepackage{droidsans}
\usepackage{charter}
\usepackage[T1]{fontenc}
\usepackage[usenames,dvipsnames]{xcolor}
\usepackage{setspace}
\usepackage[compact]{titlesec}
\usepackage{hyperref}

\usepackage{subfigure}
\usepackage{epstopdf}%

\definecolor{cream}{RGB}{222,217,201}

\newcommand{\SDiO}{SO$_\text{2}$ }
\newcommand{\STriO}{SO$_\text{3}$ }
\newcommand{\SOx}{SO$_\text{x}$ }

\newcommand{\reffig}[1]{{Fig.~\ref{#1}}}
\newcommand{\refsec}[1]{{Sec.~\ref{#1}}}

\DeclareUnicodeCharacter{2009}{\,} 

\hypersetup{hyperindex,breaklinks,colorlinks=true,linkcolor=blue,citecolor=blue,urlcolor=blue,filecolor=blue}

\begin{document}

\pagestyle{fancy}
\thispagestyle{plain}
\fancypagestyle{plain}{
\renewcommand{\headrulewidth}{0pt}
}
\makeFNbottom
\makeatletter
\renewcommand\LARGE{\@setfontsize\LARGE{15pt}{17}}
\renewcommand\Large{\@setfontsize\Large{12pt}{14}}
\renewcommand\large{\@setfontsize\large{10pt}{12}}
\renewcommand\footnotesize{\@setfontsize\footnotesize{7pt}{10}}
\makeatother

\renewcommand{\thefootnote}{\fnsymbol{footnote}}
\renewcommand\footnoterule{\vspace*{1pt}%
\color{cream}\hrule width 3.5in height 0.4pt \color{black}\vspace*{5pt}} 
\setcounter{secnumdepth}{5}

\makeatletter 
\renewcommand\@biblabel[1]{#1}            
\renewcommand\@makefntext[1]%
{\noindent\makebox[0pt][r]{\@thefnmark\,}#1}
\makeatother 
\renewcommand{\figurename}{\small{Fig.}~}
\sectionfont{\sffamily\Large}
\subsectionfont{\normalsize}
\subsubsectionfont{\bf}
\setstretch{1.125} %
\setlength{\skip\footins}{0.8cm}
\setlength{\footnotesep}{0.25cm}
\setlength{\jot}{10pt}
\titlespacing*{\section}{0pt}{4pt}{4pt}
\titlespacing*{\subsection}{0pt}{15pt}{1pt}
\fancyfoot{}
\fancyfoot[LO,RE]{\vspace{-7.1pt}\includegraphics[height=9pt]{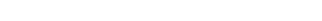}}
\fancyfoot[CO]{\vspace{-7.1pt}\hspace{13.2cm}\includegraphics{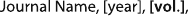}}
\fancyfoot[CE]{\vspace{-7.2pt}\hspace{-14.2cm}\includegraphics{head_foot/RF}}
\fancyfoot[RO]{\footnotesize{\sffamily{1--\pageref{LastPage} ~\textbar  \hspace{2pt}\thepage}}}
\fancyfoot[LE]{\footnotesize{\sffamily{\thepage~\textbar\hspace{3.45cm} 1--\pageref{LastPage}}}}
\fancyhead{}
\renewcommand{\headrulewidth}{0pt} 
\renewcommand{\footrulewidth}{0pt}
\setlength{\arrayrulewidth}{1pt}
\setlength{\columnsep}{6.5mm}
\setlength\bibsep{1pt}
\makeatletter 
\newlength{\figrulesep} 
\setlength{\figrulesep}{0.5\textfloatsep} 

\newcommand{\topfigrule}{\vspace*{-1pt}%
\noindent{\color{cream}\rule[-\figrulesep]{\columnwidth}{1.5pt}} }

\newcommand{\botfigrule}{\vspace*{-2pt}%
\noindent{\color{cream}\rule[\figrulesep]{\columnwidth}{1.5pt}} }

\newcommand{\dblfigrule}{\vspace*{-1pt}%
\noindent{\color{cream}\rule[-\figrulesep]{\textwidth}{1.5pt}} }

\makeatother
\twocolumn[
  \begin{@twocolumnfalse}
{\includegraphics[height=30pt]{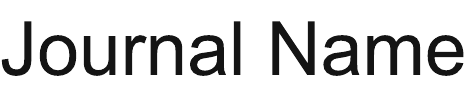}\hfill\raisebox{0pt}[0pt][0pt]{\includegraphics[height=55pt]{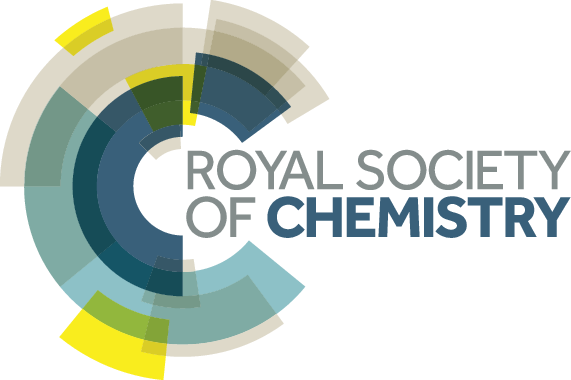}}\\[1ex]
\includegraphics[width=18.5cm]{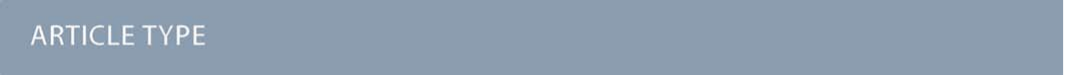}}\par
\vspace{1em}
\sffamily
\begin{tabular}{m{4.5cm} p{13.5cm} }

\includegraphics{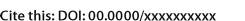} & \noindent\LARGE{\textbf{First-principles investigation of sulfur and sulfur-oxide compounds as potential optically active defects on (6,5) SWCNT}} \\
\vspace{0.3cm} & \vspace{0.3cm} \\

 & \noindent\large{Tina~N.~Mihm,\textit{$^{\P}$} Kasidet Jing Trerayapiwat,\textit{$^{\S \ddag}$} Xinxin~Li,\textit{$^{\textbar\textbar,**}$} Xuedan Ma\textit{$^{\textbar\textbar,\dag,a}$} and Sahar Sharifzadeh$^{\ast}$\textit{$^{\P,\S,b,c}$}} \\ %

\includegraphics{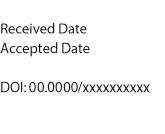} &
\noindent\normalsize{Semiconducting single-walled carbon nanotubes (SWCNT) functionalized with covalent defects are a promising class of optoelectronic materials with strong, tunable photoluminescence and demonstrated single photon emission (SPE). Here, we investigate sulfur-oxide containing compounds as a new class of optically active dopants on (6,5) SWCNT. Experimentally, it has been found that when the SWCNT is exposed to sodium dithionite, the resulting compound displays a red-shifted and bright photoluminescence peak that is characteristic of doping with covalent defects. We perform density functional theory calculations on the possible adsorbed compounds that may be the source of doping (S, SO, \SDiO and \STriO). We predict that the two smallest molecules strongly bind to the SWCNT with binding energies of $\sim 1.5-1.8$ eV and $0.56$ eV for S and SO, respectively, and introduce in-gap electronic states into the bandstructure of the tube consistent with the measured red-shift of ($0.1-0.3$) eV, consistent with measurements. In contrast, the larger compounds are found to be either unbound or weakly physisorbed with no appreciable impact on the electronic structure of the tube, indicating that they are unlikely to occur. Overall, our study suggests that sulfur-based compounds are promising new dopants for (6,5) SWCNT with tunable electronic properties.} \\

\end{tabular}

 \end{@twocolumnfalse} \vspace{0.6cm}
]
\renewcommand*\rmdefault{bch}\normalfont\upshape
\rmfamily
\section*{}
\vspace{-1cm}

\footnotetext{\textit{$^{\ast}$~EMAIL: ssharifz@bu.edu}}
\footnotetext{\textit{$^{\P}$~Department of Electrical and Computer Engineering, Boston University,  MA, 02215, USA}}
\footnotetext{\textit{$^{\S}$~Department of Chemistry, Boston University, Boston, MA, 02215, USA}}
\footnotetext{\textit{$^{\textbar\textbar}$~Center for Nanoscale Materials, Argonne National Laboratory, Lemont, IL 60439, USA}}
\footnotetext{\textit{$^{**}$~Consortium for Advanced Science and Engineering, University of Chicago, Chicago, IL 60637, USA}}
\footnotetext{\textit{$^{\dag}$~Materials Science and NanoEngineering, Rice University, Houston, TX, 77251, USA}}
\footnotetext{\textit{$^{a}$~Northwestern Argonne Institute of Science and Engineering, Evanston, IL 60208, USA}}
\footnotetext{\textit{$^{b}$~Materials Science Division, Boston University, Boston, MA, 02215, USA}}
\footnotetext{\textit{$^{c}$~Department of Physics, Boston University, Boston, MA, 02215, USA}}
\footnotetext{\textit{$^{\ddag}$~Present address: Center for Nanoscale Materials, Argonne National Laboratory, Lemont, IL 60439, USA}}

\footnotetext{\dag~Supplementary Information available: Measured spectrum of the doped SWCNT, additional binding information on the \SOx adsorbates including unbound structures, examples of the binding configurations for the two thioether-l positions,  and a table of all bond lengths and bond angles for each adsorbate. We also provide charge density differences and a graph of the band splitting across k-points for the five defects shown in Fig 4 in the text, along with all band structures and projected density of states for all adsorbates studied. Finally, We provide a comparison of our new covalent defects against previously studied covalent defects, a comparison of PBE results vs HSE06 and GW/BSE approximation for pristine and one dopant, and k-point convergence testing results. See DOI: 00.0000/00000000.}

\section{Introduction}
 Semiconducting single walled carbon nanotubes (SWCNT) doped with covalent defects display high purity single photon emission (SPE) with high quantum yields,\cite{piao_brightening_2013, ma_room-temperature_2015,wang_bright_2013, he_intrinsic_2019, miyauchi_brightening_2013} and show promise in technologies such as
sensing\cite{barone_near-infrared_2005, settele_ratiometric_2024, galonska_guanine_2023, metternich_near-infrared_2023, spreinat_quantum_2021, basu_role_2024, basu_ratiometric_2024, lin_creating_2019, spreinat_quantum_2021, mann_quantum_2020} and quantum information science.\cite{hogele_photon_2008, he_carbon_2018}
The controlled sensitivity of the SWCNT to adsorbates allows for a variety of covalent dopants to be introduced.\cite{barone_near-infrared_2005, wang_bright_2013, ma_room-temperature_2015, weight_coupling_2021, zaumseil_luminescent_2022} These dopants can induce new associated optical transitions observed as a red-shift in the optical spectrum,\cite{wang_excitons_2020, ghosh_oxygen_2010, piao_brightening_2013, miyauchi_brightening_2013, he_tunable_2017, ma_room-temperature_2015, weight_coupling_2021, he_intrinsic_2019, zaumseil_luminescent_2022} and improved SPE.\cite{lohmann_sp3-functionalization_2020, ma_electronic_2014, ghosh_oxygen_2010, he_tunable_2017, ishii_enhanced_2018, wang_bright_2013, he_carbon_2018, zaumseil_luminescent_2022} due to either defect-induced symmetry breaking~\cite{ghosh_oxygen_2010} or the presence of deep in-gap states~\cite{trerayapiwat_broken_2024}. As such, there is great interest in the discovery of new covalent defects that can be synthesized on SWCNT.

The most common covalent defects that have demonstrated SPE are atomic hydrogen\cite{kilina_brightening_2012} and oxygen,\cite{ ma_electronic_2014, ghosh_oxygen_2010, miyauchi_brightening_2013, ma_room-temperature_2015, basu_role_2024, basu_ratiometric_2024, lin_creating_2019} and functionalized aryl molecules.~\cite{he_tunable_2017, hartmann_photoluminescence_2015, ishii_enhanced_2018, piao_brightening_2013, weight_interplay_2021, setaro_preserving_2017, hayashi_azide_2022, spreinat_quantum_2021, mann_quantum_2020} Both atomic and molecular defects result in a high intensity red-shift in the tube's photoluminescence. The aryl groups have the advantage of being more tunable through their functional groups,\cite{he_tunable_2017, hartmann_photoluminescence_2015, ishii_enhanced_2018, piao_brightening_2013, he_carbon_2018}
but require careful control of the defect density to avoid excessive functionalization that results in exciton quenching.\cite{sander_controlled_2024, karousis_current_2010, wang_bright_2013}
Atomic defects are desirable because they are abundant and form at stable locations along the tube,\cite{ ghosh_oxygen_2010, miyauchi_brightening_2013, ma_room-temperature_2015, ma_electronic_2014} and, thus, may provide a practical path for industrial scale production of doped SWCNT. However, the degree of emission enhancement is generally lower for atomic defects compared to the aryl groups.~\cite{wang_bright_2013}

Regardless of the defect type, SPE and optical properties are highly-dependent on the defect binding configuration.~\cite{zaumseil_luminescent_2022,gifford_exciton_2018, ghosh_oxygen_2010} Covalent defects are typically introduced to the SWCNT through breaking of one carbon sp$^2$ bond and attaching a ligand that creates an sp$^3$ defect site. For certain defects and adsorption sites, the defect forms an sp$^2$ bond to two carbon atoms, breaking a double bond but maintaining the sp$^2$ hybridization of the tube. The photoluminescence (PL) and photochromic properties of the tube strongly depend on the location of binding and the type of bond formed,~\cite{gifford_exciton_2018, weight_coupling_2021, settele_synthetic_2021} presumably due to the varying degrees of perturbation to the SWCNT electronic structure. 

  In this article, we investigate sulfur and sulfur oxide (\SOx)-based sp$^2$ and sp$^3$ dopants as a class of atomic and small molecular covalent dopants that may be synthesized on (6,5) SWCNT. The (6,5) SWCNT is a chiral and semiconducting tube with clearly defined two lowest-energy optical transitions, E$_{11}$ and E$_{22}$ at $\sim 1.25$ eV and $\sim 2.21$ eV, respectively, with multiple optically-active dopants successfully introduced, including the oxygen atom.\cite{ ma_electronic_2014, ghosh_oxygen_2010, miyauchi_brightening_2013, ma_room-temperature_2015, basu_role_2024, basu_ratiometric_2024} In our study, sulfur and sulfur oxide-based defects were chosen as potential optically active defects because sulfur is isovalent to oxygen; thus, we expect similar bonding and optical properties associated with the sulfur doped SWCNT.
 Furthermore, when used to dope carbon materials, sulfur-containing compounds have shown efficient fluorescence quantum yield with tunable emissions,\cite{liu_novel_2023, luo_comparative_2023, ma_s-doped_2022, jin_sulfur-doped_2025}
 indicating that sulfur may be an optically-active defect on (6,5) SWCNT. Experimentally, it was shown that introduction of sodium dithionite to solution-phase (6,5) SWCNT, which is expected to break down into \SOx (predominantly \SDiO) on the surface, leads to a new doped species with the desired red-shifted strong emission. \cite{li_near-infrared_2025}
Within density functional theory (DFT), we perform a comprehensive study of the most probable configurations of adsorbed \SOx that may arise (S, SO, \SDiO, \STriO) in order to determine the most likely species that has been synthesized. We predict that the larger two molecules studied, \SDiO and \STriO, are weakly physisorbed to the SWCNT, with little to no electronic interaction to the tube, in agreement with prior DFT-based studies of these  compounds on SWCNT.\cite{chen_curvature_2017, yu_single-walled_2007, mittal_carbon_2014, yao_humidity-assisted_2011, goldoni_single-wall_2003, peymani_functionalization_2016, oftadeh_sulfur_2014, shen_dependence_2014} In contrast, S and SO favorably and strongly bind to the tube with a red-shifted gap due to either defect-induced band splitting or introduction of an in-gap defect state. S in an sp$^2$ bonding configuration is predicted to be the most energetically stable configuration. Overall, our study indicates that S and SO defects can be incorporated into and tune the optical properties of (6,5) SWCNT.

The remainder of the paper is organized as follows: \refsec{Comp-methods} outlines the DFT computational details and details of the binding configurations considered; \refsec{Results-Dis} presents the predicted adsorbate binding energies, optimized structures, and electronic structure; \refsec{Conclusions} concludes with a statement on which {\SOx} defects are possibly the optically active defects that have been observed experimentally.

\section{Computational Methods}
\label{Comp-methods}
The electronic structure and adsorption energy of the pristine and doped SWCNT were computed within density functional theory (DFT) as implemented in the Vienna Ab initio Simulation Package (VASP),~\cite{kresse_efficient_1996, kresse_efficiency_1996, kresse_ab_1993, kresse_ab_1994, kresse_norm-conserving_1994} with frozen-core projector-augmented wave (PAW) pseudopotentials describing the nuclei and core electrons.~\cite{blochl_projector_1994, kresse_ultrasoft_1999}  
The exchange-correlation was treated within the generalized gradient approximation (GGA) of Perdew, Burke, and Ernzerhof (PBE),~\cite{perdew_generalized_1996} with the addition of Grimme D3 van der Waals corrections.~\cite{grimme_consistent_2010} 
The choice of the PBE functional was made because empirical van der Waals corrections are better suited to PBE, which tends to underbind, rather than LDA, which tends to overbind.\cite{he_accuracy_2014, haas_calculation_2009} We note that while GGA underestimates the band gap, our prior studies of aryl-doped (6,5) SWCNT found that while many-body corrections open up the gap, the relative energy between dopant and tube frontier states is consistent between LDA/GGA and many-body perturbation theory~\cite{trerayapiwat_broken_2024}. Tests performed using the hybrid HSE06 functional (see SI, section 5)\cite{krukau_influence_2006} indicated that HSE06 and PBE predict the same trend in binding energies as well as band energies, validating our choice to use the less computationally costly PBE functional.

\begin{figure}
{
\centering
\subfigure[\mbox{}]{%
\includegraphics[width=0.45\textwidth,height=\textheight,keepaspectratio]{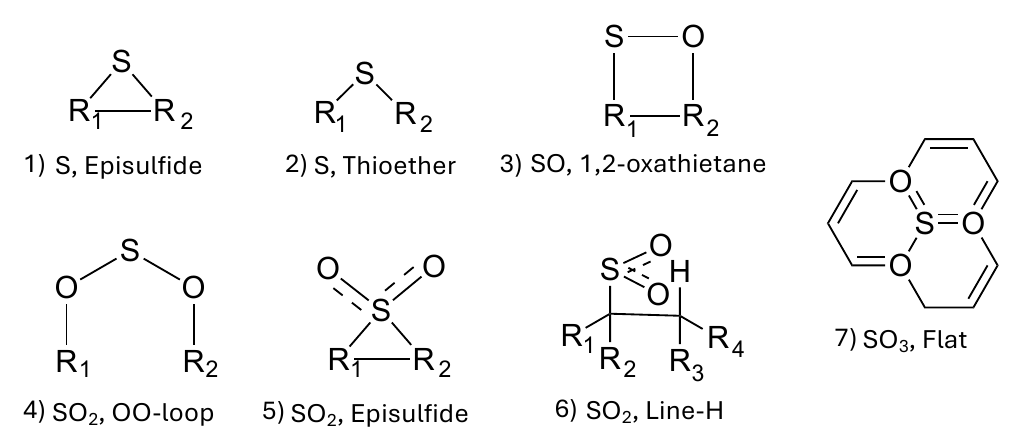}
\label{subfig1a:Structures}
}
}

{
\centering
\subfigure[\mbox{}]{%
\includegraphics[width=0.3\textwidth,height=\textheight,keepaspectratio]{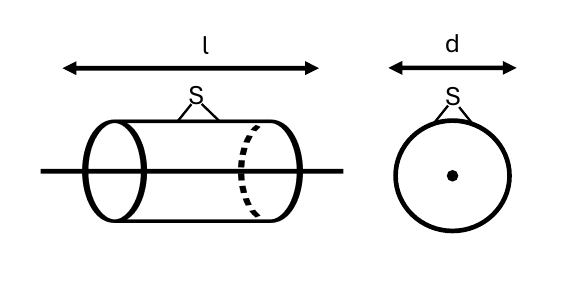}
\label{subfig1b:Orientation}
}
}
\caption{a) The configuration of different sulfur compounds studied relative to the carbon atoms of (6,5) SWCNT (labeled R$_x$). b) Illustration of the two different orientations of the sulfur compounds relative to the SWCNT tube axis with a single sulfur atom shown as an example. The cylinder represents the (6,5) SWCNT.
}
\label{Fig1:Ligand_orientation}
\end{figure}

All SWCNT were placed in a periodic cell with greater than $10$ {\AA} of vacuum along the two aperiodic directions. Restricted DFT calculations were performed for all but the passivated `Line-H' \SDiO system (See \reffig{subfig1a:Structures}), which requires spin-polarized unrestricted DFT. Calculations were performed with a $\Gamma$-centered $1\times 1\times 2$ k-point mesh, which converged the total energy to less than 1 meV/atom (see SI, section 6), and a plane wave energy cutoff of $400$ eV, which is default for VASP. The structure of the full tube and defect were optimized until all forces were less than $0.01$ eV/{\AA}. 
The optimized lattice vector of the pristine (6,5) SWCNT is $\alpha = 32.4$ {\AA}, $b = 20.0$ {\AA}, and $c = 40.4$ {\AA} and was kept constant in the periodic direction ($c$) for all the defective systems. 

The adsorbate binding energy was computed as 

\begin{equation}
\label{Equ:Binding_E}
    E_\mathrm{b} = E_\mathrm{CNT+SOx} - (E_\mathrm{CNT} + E_\mathrm{SOx}).
\end{equation}
Here, the $E_\mathrm{CNT+SOx}$ is the total energy of the adsorbate on SWCNT, $E_\mathrm{CNT}$ is the total energy of the pristine tube, and $E_\mathrm{SOx}$ is the total energy of the sulfur compound in vacuum. To obtain the total energy of each system individually, all three were placed in a box of the same size with the same computational parameters applied with the exception of the k-mesh for the isolated \SOx, which was $1 \times 1 \times 1$. For the hydrogen passivated {\SDiO} structure (structure $6$ in \reffig{subfig1a:Structures}), we subtracted out the binding energy associated with the hydrogen atom ($-1.41$ eV), which did not interact with \SDiO, from the total binding energy ($E_\mathrm{b}^{SO_2+H} = -1.74$ eV). Thus, $E_\mathrm{b}^{SO_2} = -1.74 $eV$ + 1.41$ eV = -0.31 eV.

We note that the experimental measurements were taken with the SWCNT suspended in solution of sodium dodecyl sulfate (SDS), which we do not account for in our simulations. These long carbon-chain surfactants have been shown to interact with aryl diazonium defects, with a noticeable impact on the selectivity of the dopant's binding configuration.\cite{he_low-temperature_2017, hilmer_role_2012} However, we expect this impact will be less significant for our atomic-scale defects. Exclusion of the surfactant is consistent with previous computational studies of oxygen and SO$_2$/\STriO defects on SWCNT.\cite{chen_curvature_2017, yu_single-walled_2007, mittal_carbon_2014, yao_humidity-assisted_2011, goldoni_single-wall_2003, peymani_functionalization_2016, oftadeh_sulfur_2014, shen_dependence_2014, ma_electronic_2014} In addition, because the doped nanotubes are washed prior to analysis and the \SOx molecules are closed shell, we do not expect any charging of the dopant.

\textit{Structures studied:} 
 The different sulfur compounds considered and their corresponding binding configuration relative to the SWCNT are presented in Figure \ref{subfig1a:Structures}. We study S, SO, \SDiO, and \STriO  because it was previously determined that \SDiO may decompose into these compounds on SWCNT.\cite{goldoni_spectroscopic_2004} 
 We considered different possible configurations of the adsorbate and only show configurations with favorable binding here; see supplementary materials Section 2.1 for details on unbound structures.
 The three kinds of covalent binding sites on the SWCNT are the sidewalls (side of tube), edge sites (end of tube), or near defects (such as vacancies and Stone-Wales complexes)
Consistent with prior studies,\cite{ ma_electronic_2014, ghosh_oxygen_2010, miyauchi_brightening_2013, ma_room-temperature_2015, basu_role_2024, basu_ratiometric_2024} we considered only the sidewall because the edges make up a fraction of the surface area of the tube and would not result in the observed bright luminescence peak. 
We did not consider Stone-Wales defects because there was no evidence of the presence of these defects in experiment.\cite{yu_excitons_2023, biktagirov_topological_2025}
In order to test the impact of vacancies on the binding affinity, we studied the adsorption of the SO$_2$, Line-H system in the vicinity of a single vacancy. The presence of the vacancy was found to cause SO$_2$ to become unbound, indicating that the vacancy does not enhance binding.

 For the single sulfur atom, we considered epoxide-type and ether-type binding (structures $1$ and $2$), both of which have been shown to be energetically favorable locations for oxygen binding on (6,5) SWCNT.~\cite{ma_electronic_2014} In thioether-type bonding the S atom is arranged between a broken carbon-carbon bond resulting in sp$^2$ hybridization, while for episfulfide-type, the S atom sits between a single carbon-carbon bond with sp$^3$ hybridization. For the SO molecule, a four-member ring-type binding (structure $3$) with the S and O each bound to one carbon atom was found to be favorable. For each S and SO configuration, we consider two different orientations of the C-S-C bonds, along the long (`l') and short (`d') axes of the tube (see \reffig{subfig1b:Orientation}). We also considered two different `l' placements for the thioether S structure, which are inequivalent due to the chiral symmetry of the (6,5) SWCNT: in-line with or offset from the tube's chirality (see Section 2.2 of the Supplementary materials for structures). 
For \SDiO, three different binding configurations were found to be favorable: 1) the \SDiO attached to two carbons across a broken carbon bond via the two oxygen on the \SDiO (structure $4$), which was inspired by the orientation of \SDiO bound to graphene sheets and we label ``OO-loop'' ;\cite{humeres_kinetics_2012, humeres_reactive_2017} 2) the episulfide-type binding, a modification of the epoxide-type binding of S (structure $5$); and 3) the \SDiO oriented such that the oxygen atoms are in line with a hydrogen atom that is attached to a carbon atom neighboring the adsorption site, labeled ``Line-H"  (structure $6$). The latter structure was introduced to test whether H can increase the affinity of the adsorbate to the tube. 
Lastly, for \STriO, a flat-type binding motivated by binding of \STriO on graphene is found to be favorable (structure $7$).\cite{shokuhi_rad_application_2016} 

\section{Results and Discussion}
\label{Results-Dis}
We investigate the possibility of doping (6,5) SWCNT with SO$_x$-based compounds by describing the binding and electronic structure of new compounds from first-principles calculations. The DFT-predicted binding energies of the sulfur-based compounds of Figure~\ref{Fig1:Ligand_orientation} are presented in Table \ref{Table1:CNT_binding_E}. 
The S atom is the most strongly bound to the tube, with binding energy ranging from $-1.53$ eV to $-1.82$ eV depending on the bonding arrangement. This value is greater than half of a C-S covalent bond energy of $\sim -2.7$ eV,\cite{noauthor_covalent_nodate, noauthor_fundamentals_2013} consistent with strong chemical bonding between the tube and S. The thioether-d configuration, which is the only structure studied that allowed the tube to retain its sp$^2$ hybridization, is the most strongly bound. Additionally, the binding is stronger along the short axis (`d') than along the long axis (`l') by $0.25 - 0.29$ eV for the thioether and $0.12 - 0.16$ eV for episulfide. This is consistent with previous studies of oxygen dopants on (6,5) SWCNT (see Table S.2 in the SI for comparison).~\cite{ma_electronic_2014} We note that the binding is slightly stronger for the ``l'' position that is aligned with the tube's chirality compared with that off-line structure by 50 meV. Overall, these findings indicate that the binding of the sulfur on the CNT is influenced by its orientation, presumably due to the chiral nature of the tube. 

\begin{table}[]
\caption{Calculated binding energy ($E_\mathrm{b}$), the change in the energy difference between highest occupied and lowest unoccupied state upon adsorption ($\Delta E_\mathrm{gap}$), and shift in Fermi energy with respect to the pristine system ($\Delta E_\mathrm{f}$) for the defective (6,5) SWCNT configurations considered in this work. $E_\mathrm{b}$ is defined such that a negative energy is bound (see Eq \ref{Equ:Binding_E}). All energies are in eV.}
\label{Table1:CNT_binding_E}
\vspace{10pt}
\resizebox{0.95\columnwidth}{!}{%
\begin{tabular}{lccc}
\hline
Structure         & $E_\mathrm{b}$ & $\Delta E_\mathrm{gap}$ & $\Delta E_\mathrm{f}$ \\
\hline
S, Episulfide-d       & -1.69                  & 0.04         & -0.04             \\
S, Thioether-d        & -1.82                 & 0.02          & -0.01             \\
S, Episulfide-l \\ Off-set with chiral direction       & -1.53                 & 0.28          & 0.05              \\
S, Episulfide-l \\ In-line with chiral direction        & -1.57                 & 0.05          & -0.03              \\
SO, 1,2-oxathietane-l & -0.56                  & 0.08          & -0.07             \\
SO,1,2-oxathietane-d  & -0.56                  & 0.08          & -0.07             \\
SO2, OO-loop-d        & -0.20                   & 0.00          & -0.04             \\
SO2, Episulfide-d     & -0.20                   & 0.00          & -0.05             \\
SO2, Line-H           & -0.33                 & 0.26          & 0.17              \\
SO3, flat             & -0.27                & 0.00          & -0.08             \\
Pristine              & --                   & 0.00          & --              \\
\hline
Experimental peak shift & --& 0.12 - 0.26  &--\\
\hline
\end{tabular}
}
\end{table}

For the \SOx molecules, binding is much weaker with binding energies $<0.6$ eV for all molecular configurations studied. The binding energy of the SO molecule is strongest at $-0.56$ eV for binding along both the long and short axes. \SDiO and \STriO display much weaker binding to the surface, more consistent with physisorption. For \SDiO, all configurations studied were predicted to bind via physisorption with energies ranging from $-0.20$ eV (no hydrogen) to $-0.33$ eV (with hydrogen), indicating that the presence of hydrogen only slightly enhances the bond strength. For \STriO, the binding energy is similarly weak at $-0.27$ eV. That the larger two defects, \SDiO and \STriO, are physisorbed on the surface is consistent with previous studies on carbon nanotubes.~\cite{chen_curvature_2017, yu_single-walled_2007, mittal_carbon_2014, yao_humidity-assisted_2011, goldoni_single-wall_2003, peymani_functionalization_2016, oftadeh_sulfur_2014}
The weaker binding for the 
SO$_x$ molecules compared with a single sulfur atom can possibly be attributed to the fact that S-O bonds are already quite strong and stabilize the SO$_x$ molecule, decreasing the molecule's affinity for forming new bonds.
Additionally, the presence of oxygen makes the molecules larger and more bulky than with a single sulfur atom, with steric hindrance leading to a weaker SO$_x$ bond compared to the sulfur atom.

To better understand the binding of \SOx to (6,5) SWCNT, we present the bonding arrangements of select compounds including bond lengths in  Figure \ref{fig:Structure_bonds} (see Supplementary materials Table S.1 for all compounds). For the atomic S adsorbate, the bond length is $1.76 - 1.88$ {\AA}, close to the covalent bond length of $\sim 1.8$ {\AA}.~\cite{haynes_crc_2016} The bond angle between S and the two bonding carbons on tube (R$_1$ and R$_2$ in Fig \ref{subfig1a:Structures}) are $50.2^\circ - 66.4^\circ$ (depending on orientation). The S-C bond is slightly stretched for SO adsorption ($\sim2.05$ {\AA}) with the bond angles indicating a tilt towards one of the two carbons on the tube (S-R-R angle: $\sim66^\circ$ and $\sim71^\circ$), which we attribute to steric effects due to the presence of the oxygen atom.
The \SDiO and the \STriO structures all have a predicted large adsorption distance of $\sim 3.0$ {\AA} away from the SWCNT, consistent with their weak binding. For \SDiO, the bond angles for the OO-loop-d (S-R-R angle: $82.4^\circ$ and $72.3^\circ$) and episulfide-d (S-R-R angle: $78.3^\circ$ and $75.9^\circ$) structures indicate the molecule adsorbs slightly off of the C-C bond with S in the hollow site, while the hydrogen passivated structure shows the S in the \SDiO adsorbs on top of a carbon atom (S-R-R angle: $61.2^\circ$ and $93.8^\circ$).
The slight offset of the un-passivated SO$_2$ molecule from the carbon atoms may be attributed to steric effects between the SO$_2$ dipole and the SWCNT surface. 
\STriO bond angles (S-R-R angle: $57.5^\circ$ and $94.0^\circ$) also show adsorption of S on top of one of the carbons. 

\begin{figure*}
\centering
\includegraphics[width=0.75\textwidth,height=\textheight,keepaspectratio]{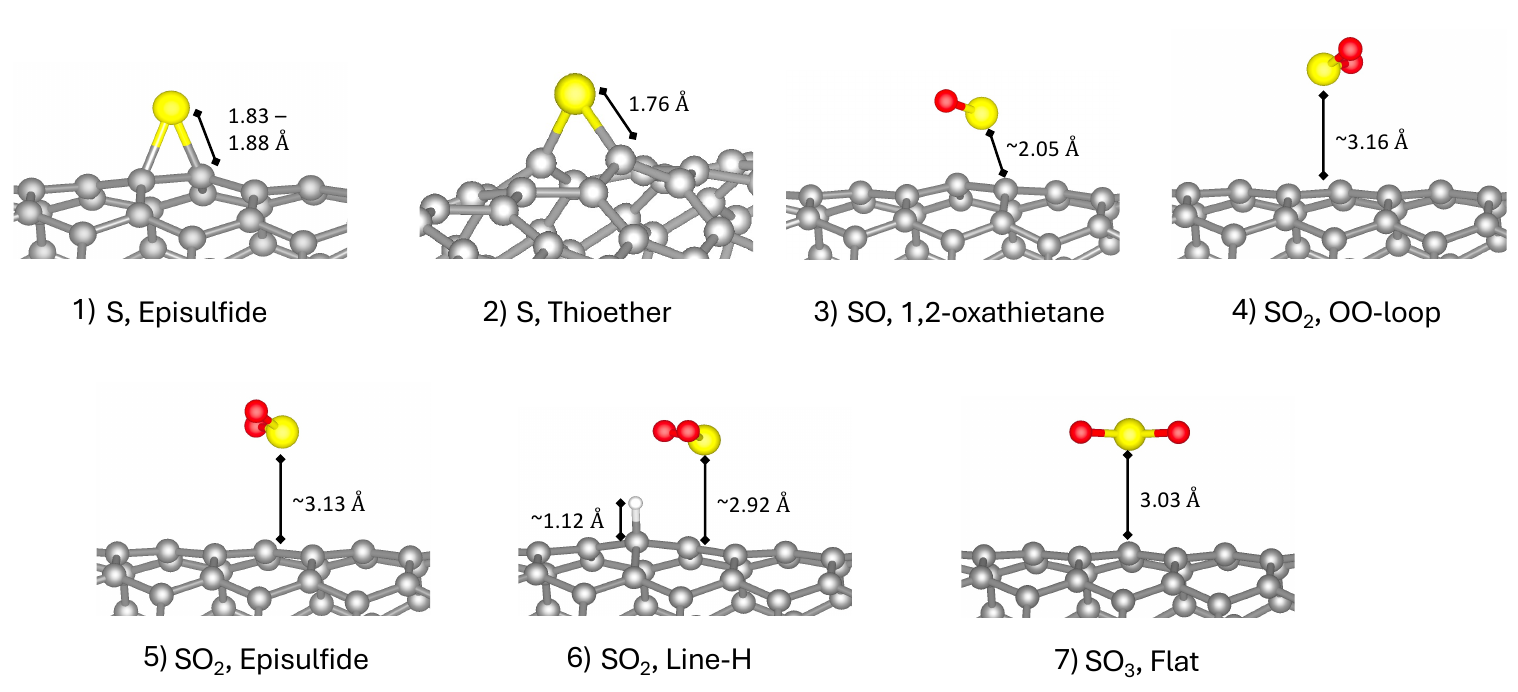}
\caption{The predicted adsorption geometry of \SOx derivatives on (6,5) SWCNT within PBE-D3. Relevant bond lengths are labeled.    
}
\label{fig:Structure_bonds}
\end{figure*}

Next, we consider the impact of S, SO, \SDiO, and \STriO on the electronic structure of (6,5) SWCNT. Based on the predicted binding for the four compounds, we may expect that the S and SO defects will impact the bandstructure significantly due to their strong bond with the tube. 
Table \ref{Table1:CNT_binding_E} presents the energy gap between occupied and unoccupied states for all bound structures studies. The gap of the doped structures ranges from 0.64 eV to 0.92 eV, compared with the pristine tube gap of 0.92 eV. The strongly bound S and SO adsorbate always result in a reduced gap with respect to the pristine tube by $20 - 300$ meV, indicating that the optical transitions will be red-shifted as a result of doping. This is in good agreement with experimental finding that the tube photoluminescence peak shifts by $100-300$ meV upon exposure to sodium dithionate.\cite{li_near-infrared_2025} Table \ref{Table1:CNT_binding_E} also shows the shift in the Fermi energy of the tube upon doping. For all but two structures (with sulfur atoms in the `l' orientation), the shift is negative, indicating a slight negative charge transfer to the surface. However, the magnitude of all the differences is small, indicating weak doping.

\begin{figure*}
\centering
\includegraphics[width=0.95\textwidth,height=\textheight,keepaspectratio]{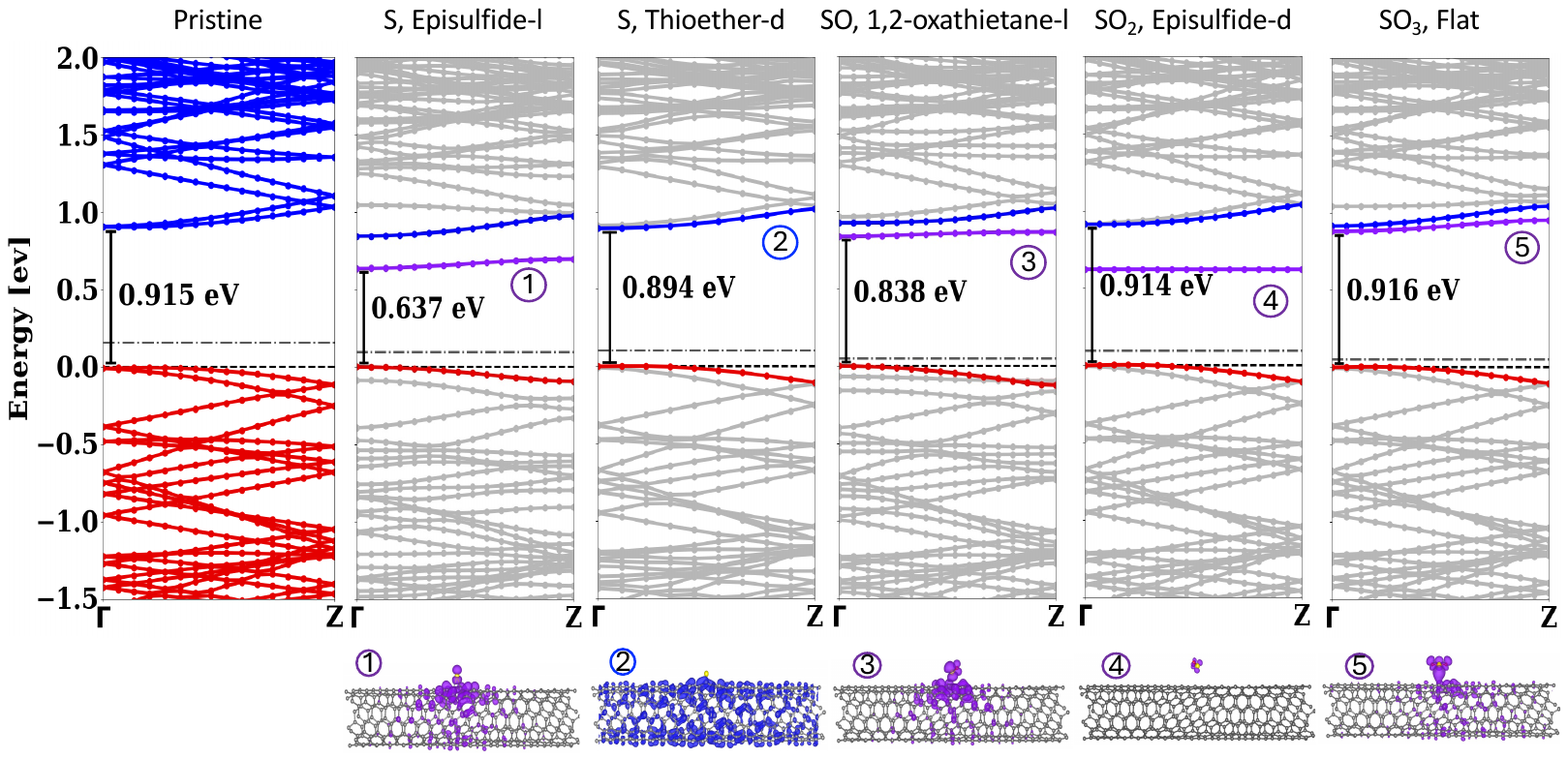}

\caption{PBE-predicted bandstructure for pristine (6,5) SWCNT and (6,5) SWCNT with the mostly likely adsorbate structures for S, SO, \SDiO, and \STriO. Occupied pristine-like valence bands and unoccupied conduction bands are shown in red and blue, respectively, while unoccupied states localized on the defect are in purple. The gray dash-dot line on each band structure represents the Fermi energy. All plots are shifted such that the top of the pristine-like valence band is at zero at $\mathbf{k} = 0$ (indicated by black dash line). The orbital charge density for the lowest energy unoccupied band is shown below the plots with an isosurface that captures $40\%$ of the density.}
\label{fig:BS_comparison}
\end{figure*}

Representative bandstructures for the tube doped with each defect type are shown in Figure~\ref{fig:BS_comparison} with all bandstructures presented in the supplementary materials Figure S6. The near-gap states of the (6,5) SWCNT are of $\pi$-type character with a two-fold degeneracy at $\mathbf{k} = 0$ due to the underlying graphitic structure. By inspection of the predicted bandstructures, we conclude that all of the defects studied break the symmetry of the tube, leading to splitting of the nearly-degenerate valence and conduction bands, consistent with previous reports for doped (6,5) SWCNT~\cite{hartmann_photoluminescence_2015, miyauchi_brightening_2013, kilina_brightening_2012, ghosh_oxygen_2010, ma_electronic_2014, piao_brightening_2013, trerayapiwat_broken_2024, weight_coupling_2021} This splitting is largest for S and SO (see Supplementary Materials Figure S.5 for more details). 
The maximum splitting within the conduction (valence) bands for episulfide-l S, thioether-d S, and SO are 200 meV (110 meV), 19 meV (9 meV) and 38 meV (67 meV), respectively. As noted previously for the oxygen dopant bound to SWCNT,\cite{ghosh_oxygen_2010} the ether-type bond maintains the sp$^2$ character of the tube and so the perturbation to the system is weaker than epoxide-type bonding, which creates an sp$^3$ bond on the tube. This weaker perturbation results in much smaller band splitting. The splitting of the bands for \SDiO is less than 4 meV for both valence and conduction bands, while for \STriO it is 4 meV for the valence and 0.13 eV for the conduction band, consistent with the weaker interaction with the tube. 

In addition to symmetry breaking-induced band splitting, both S and SO oriented along the long axis of the tube (`l' orientation) introduce a new unoccupied in-gap state associated with the defect into bandstructure, reducing the gap by up to 0.28 eV for S and 0.08 eV for SO. This in-gap state is associated with a localized defect-centered orbital hybridized with the SWCNT bands as shown in the orbital density plots below the bandstructure. This state is not present for S in the thioether-d configuration for which no defect-centered orbital is present near the gap. Interestingly, whether the defect is in-line with the chiral direction of the tube plays a role in the location of the in-gap state for the `l'-oriented S defects: if the position of S is aligned with the tube's chirality the defect-centered state is significantly closer in energy to the conduction band. 

\SDiO also introduces an in-gap state with an associated orbital density that is localized on the defect, indicating no interaction with the tube. For \STriO, there is an in-gap state resonant with the conduction bands. This state is slightly delocalized over parts of the tube and slightly dispersive. However, we suspect that this mixing between the defect state and conduction band may be artificial and due to DFT self-interaction error which results in mixing of two resonant states, as occurs in bond dissociation.~\cite{ruzsinszky_spurious_2006} 

Considering both the predicted binding energy and the change in the bandstructure upon adsorption, we expect that only S and SO are candidate covalent dopants to (6,5) SWCNT consistent with experiment with the most likely being a combination of the three sulfur atom configurations and a small amount of the SO dopants.
These defects chemically bind to the SWCNT and result in in-gap states that decrease the band gap by 0.04 - 0.28 eV, consistent in energy with the red-shifted peaks in experiment at 0.12 - 0.26 eV (see Table~\ref{Table1:CNT_binding_E}). These transitions may be $\pi-\pi*$ excitations from symmetry broken band-edges configuration (as for the S atom in thioether-d) or may involve defect-localized in-gap unoccupied states (as for S in thioether-l configuration). We do not expect SO$_2$ or SO$_3$ to bind to the tube based on their weak adsorption energy. The fact that these larger molecules are physisorbed and do not impact the bandgap of the tube indicates that, even if they do bind, they do not contribute to the red-shifted PL.

Table S.2 compares the predicted binding energy and band gap reduction for our S/SO$_x$ defects and other previously studied covalent defects. We note that compared to oxygen, for which the adsorption energy varies by $> 1$ eV depending on the adsorption site (see Table S.2), the sulfur binding sites are quite close in energy. Thus, we expect sulfur to bind across the different binding sites of the SWCNT, which we predict will have differing electronic transitions from their associated banstructure shown in Figure~\ref{fig:BS_comparison}. This prediction is consistent with the increased structure and broadening in the PL spectrum of sulfur-doped SWCNT compared with the oxygen-doped tube.

Lastly, we note that while the exact binding energies and red-shift of the gap depends on the level of theory considered, we expect the trends to be consistent among different levels of theory, as evidenced by select calculations presented in Table S.3.

Overall, we determine that the binding of the dopant to the SWCNT disrupts the carbon sp2 bonds, with the perturbation causing the red-shift seen in the PL spectrum. For the large defects that physisorb on the tube, the disruption is minimal, while for the chemically bound SO and S atoms, the impacts is significant and observable.
 
\section{Conclusions}
\label{Conclusions}

In summary, DFT calculations were used to understand the binding of sulfur-based covalent defects on (6,5) SWCNT. We performed DFT calculations of the adsorption energy and bandstructure of four possible \SOx defects (S, SO, \SDiO, and \STriO) onto (6,5) SWCNT. We determined that the two smaller defects, S and SO are strongly bound, with bond distances consistent with a carbon-sulfur covalent bond. 
In addition, there is a strong dependence of binding energy on the adsorption location along the tube. Furthermore, we predict S in the sp$^2$ hybridized thioether-d position is the most strongly bound at $1.8$ eV and therefore most likely to form, reducing the gap by $\sim 20$ meV, slightly smaller than the red-shift of the bright peak in experiment. S in the sp$^3$ hybridized episulfide-l is also strongly bound and red-shifts the bandgap by up to $0.3$ eV, consistent with the observed low-energy shoulder in the PL spectrum. SO is also strongly bound to the CNT by 0.56 eV and red-shifts the gap by 80 meV, consistent with the brightest peak in the PL spectrum. Our analysis indicates that both S and SO have potential for use as an emissive defect for use in tuning the photoluminescence energy of the SWCNT and may be able to form through the synthetic technique given in Ref \cite{li_near-infrared_2025}. 

\section{Author Contributions}

{\bf{Tina~N.~Mihm}} - Conceptualization (equal); Formal Analysis (lead); Investigation (lead); Visualization; Writing - Original draft preparation (equal); Writing - review and editing (equal)

{\bf{Kasidet Jing Trerayapiwat}} - Investigation (supporting); Writing - review and editing (equal); Resources

{\bf{Xinxin~Li}} - Investigation (supporting); Formal Analysis (supporting); Writing - review and editing (equal)

{\bf{Xuedan Ma}} - Conceptualization (lead); Supervision (lead); Writing - review and editing (equal)

{\bf{Sahar Sharifzadeh}} - Conceptualization (lead); Supervision (lead); Writing - Original draft preparation (equal); Writing - review and editing (equal)

\section{Conflict of Interest}
The authors have no conflicts to disclose.

\section{Data Availability}

Data for this article, including inputs and outputs for all VASP calculations are available at SOx-CNT at \url{https://github.com/fpmats/Calculation_IO/tree/main/SOx-CNT}
All other data supporting this article have been included as part of the main text and Supplementary Information.

\section{Acknowledgements}

S.S. and T.M. acknowledge financial support from the U.S. Department of Energy (DOE), Office of Science, Basic Energy Sciences under Award No. DE-SC0023402. X.L. acknowledges support from the National Science Foundation DMR Program under award no. DMR-1905990. Work performed at the Center for Nanoscale Materials, a U.S. Department of Energy Office of Science User Facility, was supported by the U.S. DOE, Office of Basic Energy Sciences, under Contract No. DE-AC02-06CH11357. We would like to acknowledge computational resources from the National Energy Research Scientific Computing Center (NERSC), a DOE Office of Science User Facility supported by the Office of Science of the U.S. Department of Energy under Contract No. DE-AC02-05CH11231, and Boston University’s Research Computing Services.

\balance

\providecommand*{\mcitethebibliography}{\thebibliography}
\csname @ifundefined\endcsname{endmcitethebibliography}
{\let\endmcitethebibliography\endthebibliography}{}


\begin{mcitethebibliography}{71}
\providecommand*{\natexlab}[1]{#1}
\providecommand*{\mciteSetBstSublistMode}[1]{}
\providecommand*{\mciteSetBstMaxWidthForm}[2]{}
\providecommand*{\mciteBstWouldAddEndPuncttrue}
  {\def\EndOfBibitem{\unskip.}}
\providecommand*{\mciteBstWouldAddEndPunctfalse}
  {\let\EndOfBibitem\relax}
\providecommand*{\mciteSetBstMidEndSepPunct}[3]{}
\providecommand*{\mciteSetBstSublistLabelBeginEnd}[3]{}
\providecommand*{\EndOfBibitem}{}
\mciteSetBstSublistMode{f}
\mciteSetBstMaxWidthForm{subitem}
{(\emph{\alph{mcitesubitemcount}})}
\mciteSetBstSublistLabelBeginEnd{\mcitemaxwidthsubitemform\space}
{\relax}{\relax}

\bibitem[Piao \emph{et~al.}(2013)Piao, Meany, Powell, Valley, Kwon, Schatz, and Wang]{piao_brightening_2013}
Y.~Piao, B.~Meany, L.~R. Powell, N.~Valley, H.~Kwon, G.~C. Schatz and Y.~Wang, \emph{Nature Chemistry}, 2013, \textbf{5}, 840--845\relax
\mciteBstWouldAddEndPuncttrue
\mciteSetBstMidEndSepPunct{\mcitedefaultmidpunct}
{\mcitedefaultendpunct}{\mcitedefaultseppunct}\relax
\EndOfBibitem
\bibitem[Ma \emph{et~al.}(2015)Ma, Hartmann, Baldwin, Doorn, and Htoon]{ma_room-temperature_2015}
X.~Ma, N.~F. Hartmann, J.~K.~S. Baldwin, S.~K. Doorn and H.~Htoon, \emph{Nature Nanotechnology}, 2015, \textbf{10}, 671--675\relax
\mciteBstWouldAddEndPuncttrue
\mciteSetBstMidEndSepPunct{\mcitedefaultmidpunct}
{\mcitedefaultendpunct}{\mcitedefaultseppunct}\relax
\EndOfBibitem
\bibitem[Wang and Strano(2013)]{wang_bright_2013}
Q.~H. Wang and M.~S. Strano, \emph{Nature Chemistry}, 2013, \textbf{5}, 812--813\relax
\mciteBstWouldAddEndPuncttrue
\mciteSetBstMidEndSepPunct{\mcitedefaultmidpunct}
{\mcitedefaultendpunct}{\mcitedefaultseppunct}\relax
\EndOfBibitem
\bibitem[He \emph{et~al.}(2019)He, Sun, Gifford, Tretiak, Piryatinski, Li, Htoon, and Doorn]{he_intrinsic_2019}
X.~He, L.~Sun, B.~J. Gifford, S.~Tretiak, A.~Piryatinski, X.~Li, H.~Htoon and S.~K. Doorn, \emph{Nanoscale}, 2019, \textbf{11}, 9125--9132\relax
\mciteBstWouldAddEndPuncttrue
\mciteSetBstMidEndSepPunct{\mcitedefaultmidpunct}
{\mcitedefaultendpunct}{\mcitedefaultseppunct}\relax
\EndOfBibitem
\bibitem[Miyauchi \emph{et~al.}(2013)Miyauchi, Iwamura, Mouri, Kawazoe, Ohtsu, and Matsuda]{miyauchi_brightening_2013}
Y.~Miyauchi, M.~Iwamura, S.~Mouri, T.~Kawazoe, M.~Ohtsu and K.~Matsuda, \emph{Nature Photonics}, 2013, \textbf{7}, 715--719\relax
\mciteBstWouldAddEndPuncttrue
\mciteSetBstMidEndSepPunct{\mcitedefaultmidpunct}
{\mcitedefaultendpunct}{\mcitedefaultseppunct}\relax
\EndOfBibitem
\bibitem[Barone \emph{et~al.}(2005)Barone, Baik, Heller, and Strano]{barone_near-infrared_2005}
P.~W. Barone, S.~Baik, D.~A. Heller and M.~S. Strano, \emph{Nature Materials}, 2005, \textbf{4}, 86--92\relax
\mciteBstWouldAddEndPuncttrue
\mciteSetBstMidEndSepPunct{\mcitedefaultmidpunct}
{\mcitedefaultendpunct}{\mcitedefaultseppunct}\relax
\EndOfBibitem
\bibitem[Settele \emph{et~al.}(2024)Settele, Schrage, Jung, Michel, Li, Flavel, Hashmi, Kruss, and Zaumseil]{settele_ratiometric_2024}
S.~Settele, C.~A. Schrage, S.~Jung, E.~Michel, H.~Li, B.~S. Flavel, A.~S.~K. Hashmi, S.~Kruss and J.~Zaumseil, \emph{Nature Communications}, 2024, \textbf{15}, 706\relax
\mciteBstWouldAddEndPuncttrue
\mciteSetBstMidEndSepPunct{\mcitedefaultmidpunct}
{\mcitedefaultendpunct}{\mcitedefaultseppunct}\relax
\EndOfBibitem
\bibitem[Galonska \emph{et~al.}(2023)Galonska, Mohr, Schrage, Schnitzler, and Kruss]{galonska_guanine_2023}
P.~Galonska, J.~M. Mohr, C.~A. Schrage, L.~Schnitzler and S.~Kruss, \emph{The Journal of Physical Chemistry Letters}, 2023, \textbf{14}, 3483--3490\relax
\mciteBstWouldAddEndPuncttrue
\mciteSetBstMidEndSepPunct{\mcitedefaultmidpunct}
{\mcitedefaultendpunct}{\mcitedefaultseppunct}\relax
\EndOfBibitem
\bibitem[Metternich \emph{et~al.}(2023)Metternich, Wartmann, Sistemich, Nißler, Herbertz, and Kruss]{metternich_near-infrared_2023}
J.~T. Metternich, J.~A.~C. Wartmann, L.~Sistemich, R.~Nißler, S.~Herbertz and S.~Kruss, \emph{Journal of the American Chemical Society}, 2023, \textbf{145}, 14776--14783\relax
\mciteBstWouldAddEndPuncttrue
\mciteSetBstMidEndSepPunct{\mcitedefaultmidpunct}
{\mcitedefaultendpunct}{\mcitedefaultseppunct}\relax
\EndOfBibitem
\bibitem[Spreinat \emph{et~al.}(2021)Spreinat, Dohmen, Lüttgens, Herrmann, Klepzig, Nißler, Weber, Mann, Lauth, and Kruss]{spreinat_quantum_2021}
A.~Spreinat, M.~M. Dohmen, J.~Lüttgens, N.~Herrmann, L.~F. Klepzig, R.~Nißler, S.~Weber, F.~A. Mann, J.~Lauth and S.~Kruss, \emph{The Journal of Physical Chemistry C}, 2021, \textbf{125}, 18341--18351\relax
\mciteBstWouldAddEndPuncttrue
\mciteSetBstMidEndSepPunct{\mcitedefaultmidpunct}
{\mcitedefaultendpunct}{\mcitedefaultseppunct}\relax
\EndOfBibitem
\bibitem[Basu \emph{et~al.}(2024)Basu, Hendler-Neumark, and Bisker]{basu_role_2024}
S.~Basu, A.~Hendler-Neumark and G.~Bisker, \emph{ACS Nano}, 2024, \textbf{18}, 34134--34146\relax
\mciteBstWouldAddEndPuncttrue
\mciteSetBstMidEndSepPunct{\mcitedefaultmidpunct}
{\mcitedefaultendpunct}{\mcitedefaultseppunct}\relax
\EndOfBibitem
\bibitem[Basu \emph{et~al.}(2024)Basu, Hendler-Neumark, and Bisker]{basu_ratiometric_2024}
S.~Basu, A.~Hendler-Neumark and G.~Bisker, \emph{The Journal of Physical Chemistry Letters}, 2024, \textbf{15}, 10425--10434\relax
\mciteBstWouldAddEndPuncttrue
\mciteSetBstMidEndSepPunct{\mcitedefaultmidpunct}
{\mcitedefaultendpunct}{\mcitedefaultseppunct}\relax
\EndOfBibitem
\bibitem[Lin \emph{et~al.}(2019)Lin, Bachilo, Zheng, Tsedev, Huang, Weisman, and Belcher]{lin_creating_2019}
C.-W. Lin, S.~M. Bachilo, Y.~Zheng, U.~Tsedev, S.~Huang, R.~B. Weisman and A.~M. Belcher, \emph{Nature Communications}, 2019, \textbf{10}, 2874\relax
\mciteBstWouldAddEndPuncttrue
\mciteSetBstMidEndSepPunct{\mcitedefaultmidpunct}
{\mcitedefaultendpunct}{\mcitedefaultseppunct}\relax
\EndOfBibitem
\bibitem[Mann \emph{et~al.}(2020)Mann, Herrmann, Opazo, and Kruss]{mann_quantum_2020}
F.~A. Mann, N.~Herrmann, F.~Opazo and S.~Kruss, \emph{Angewandte Chemie International Edition}, 2020, \textbf{59}, 17732--17738\relax
\mciteBstWouldAddEndPuncttrue
\mciteSetBstMidEndSepPunct{\mcitedefaultmidpunct}
{\mcitedefaultendpunct}{\mcitedefaultseppunct}\relax
\EndOfBibitem
\bibitem[Högele \emph{et~al.}(2008)Högele, Galland, Winger, and Imamoğlu]{hogele_photon_2008}
A.~Högele, C.~Galland, M.~Winger and A.~Imamoğlu, \emph{Physical Review Letters}, 2008, \textbf{100}, 217401\relax
\mciteBstWouldAddEndPuncttrue
\mciteSetBstMidEndSepPunct{\mcitedefaultmidpunct}
{\mcitedefaultendpunct}{\mcitedefaultseppunct}\relax
\EndOfBibitem
\bibitem[He \emph{et~al.}(2018)He, Htoon, Doorn, Pernice, Pyatkov, Krupke, Jeantet, Chassagneux, and Voisin]{he_carbon_2018}
X.~He, H.~Htoon, S.~K. Doorn, W.~H.~P. Pernice, F.~Pyatkov, R.~Krupke, A.~Jeantet, Y.~Chassagneux and C.~Voisin, \emph{Nature Materials}, 2018, \textbf{17}, 663--670\relax
\mciteBstWouldAddEndPuncttrue
\mciteSetBstMidEndSepPunct{\mcitedefaultmidpunct}
{\mcitedefaultendpunct}{\mcitedefaultseppunct}\relax
\EndOfBibitem
\bibitem[Weight \emph{et~al.}(2021)Weight, Sifain, Gifford, Kilin, Kilina, and Tretiak]{weight_coupling_2021}
B.~M. Weight, A.~E. Sifain, B.~J. Gifford, D.~Kilin, S.~Kilina and S.~Tretiak, \emph{The Journal of Physical Chemistry Letters}, 2021, \textbf{12}, 7846--7853\relax
\mciteBstWouldAddEndPuncttrue
\mciteSetBstMidEndSepPunct{\mcitedefaultmidpunct}
{\mcitedefaultendpunct}{\mcitedefaultseppunct}\relax
\EndOfBibitem
\bibitem[Zaumseil(2022)]{zaumseil_luminescent_2022}
J.~Zaumseil, \emph{Advanced Optical Materials}, 2022, \textbf{10}, 2101576\relax
\mciteBstWouldAddEndPuncttrue
\mciteSetBstMidEndSepPunct{\mcitedefaultmidpunct}
{\mcitedefaultendpunct}{\mcitedefaultseppunct}\relax
\EndOfBibitem
\bibitem[Wang and Berkelbach(2020)]{wang_excitons_2020}
X.~Wang and T.~C. Berkelbach, \emph{Journal of Chemical Theory and Computation}, 2020, \textbf{16}, 3095--3103\relax
\mciteBstWouldAddEndPuncttrue
\mciteSetBstMidEndSepPunct{\mcitedefaultmidpunct}
{\mcitedefaultendpunct}{\mcitedefaultseppunct}\relax
\EndOfBibitem
\bibitem[Ghosh \emph{et~al.}(2010)Ghosh, Bachilo, Simonette, Beckingham, and Weisman]{ghosh_oxygen_2010}
S.~Ghosh, S.~M. Bachilo, R.~A. Simonette, K.~M. Beckingham and R.~B. Weisman, \emph{Science}, 2010, \textbf{330}, 1656--1659\relax
\mciteBstWouldAddEndPuncttrue
\mciteSetBstMidEndSepPunct{\mcitedefaultmidpunct}
{\mcitedefaultendpunct}{\mcitedefaultseppunct}\relax
\EndOfBibitem
\bibitem[He \emph{et~al.}(2017)He, Hartmann, Ma, Kim, Ihly, Blackburn, Gao, Kono, Yomogida, Hirano, Tanaka, Kataura, Htoon, and Doorn]{he_tunable_2017}
X.~He, N.~F. Hartmann, X.~Ma, Y.~Kim, R.~Ihly, J.~L. Blackburn, W.~Gao, J.~Kono, Y.~Yomogida, A.~Hirano, T.~Tanaka, H.~Kataura, H.~Htoon and S.~K. Doorn, \emph{Nature Photonics}, 2017, \textbf{11}, 577--582\relax
\mciteBstWouldAddEndPuncttrue
\mciteSetBstMidEndSepPunct{\mcitedefaultmidpunct}
{\mcitedefaultendpunct}{\mcitedefaultseppunct}\relax
\EndOfBibitem
\bibitem[Lohmann \emph{et~al.}(2020)Lohmann, Trerayapiwat, Niklas, Poluektov, Sharifzadeh, and Ma]{lohmann_sp3-functionalization_2020}
S.-H. Lohmann, K.~J. Trerayapiwat, J.~Niklas, O.~G. Poluektov, S.~Sharifzadeh and X.~Ma, \emph{ACS Nano}, 2020, \textbf{14}, 17675--17682\relax
\mciteBstWouldAddEndPuncttrue
\mciteSetBstMidEndSepPunct{\mcitedefaultmidpunct}
{\mcitedefaultendpunct}{\mcitedefaultseppunct}\relax
\EndOfBibitem
\bibitem[Ma \emph{et~al.}(2014)Ma, Adamska, Yamaguchi, Yalcin, Tretiak, Doorn, and Htoon]{ma_electronic_2014}
X.~Ma, L.~Adamska, H.~Yamaguchi, S.~E. Yalcin, S.~Tretiak, S.~K. Doorn and H.~Htoon, \emph{ACS Nano}, 2014, \textbf{8}, 10782--10789\relax
\mciteBstWouldAddEndPuncttrue
\mciteSetBstMidEndSepPunct{\mcitedefaultmidpunct}
{\mcitedefaultendpunct}{\mcitedefaultseppunct}\relax
\EndOfBibitem
\bibitem[Ishii \emph{et~al.}(2018)Ishii, He, Hartmann, Machiya, Htoon, Doorn, and Kato]{ishii_enhanced_2018}
A.~Ishii, X.~He, N.~F. Hartmann, H.~Machiya, H.~Htoon, S.~K. Doorn and Y.~K. Kato, \emph{Nano Letters}, 2018, \textbf{18}, 3873--3878\relax
\mciteBstWouldAddEndPuncttrue
\mciteSetBstMidEndSepPunct{\mcitedefaultmidpunct}
{\mcitedefaultendpunct}{\mcitedefaultseppunct}\relax
\EndOfBibitem
\bibitem[Trerayapiwat \emph{et~al.}(2024)Trerayapiwat, Li, Ma, and Sharifzadeh]{trerayapiwat_broken_2024}
K.~J. Trerayapiwat, X.~Li, X.~Ma and S.~Sharifzadeh, \emph{Nano Letters}, 2024, \textbf{24}, 667--671\relax
\mciteBstWouldAddEndPuncttrue
\mciteSetBstMidEndSepPunct{\mcitedefaultmidpunct}
{\mcitedefaultendpunct}{\mcitedefaultseppunct}\relax
\EndOfBibitem
\bibitem[Kilina \emph{et~al.}(2012)Kilina, Ramirez, and Tretiak]{kilina_brightening_2012}
S.~Kilina, J.~Ramirez and S.~Tretiak, \emph{Nano Letters}, 2012, \textbf{12}, 2306--2312\relax
\mciteBstWouldAddEndPuncttrue
\mciteSetBstMidEndSepPunct{\mcitedefaultmidpunct}
{\mcitedefaultendpunct}{\mcitedefaultseppunct}\relax
\EndOfBibitem
\bibitem[Hartmann \emph{et~al.}(2015)Hartmann, Yalcin, Adamska, Hároz, Ma, Tretiak, Htoon, and Doorn]{hartmann_photoluminescence_2015}
N.~F. Hartmann, S.~E. Yalcin, L.~Adamska, E.~H. Hároz, X.~Ma, S.~Tretiak, H.~Htoon and S.~K. Doorn, \emph{Nanoscale}, 2015, \textbf{7}, 20521--20530\relax
\mciteBstWouldAddEndPuncttrue
\mciteSetBstMidEndSepPunct{\mcitedefaultmidpunct}
{\mcitedefaultendpunct}{\mcitedefaultseppunct}\relax
\EndOfBibitem
\bibitem[Weight \emph{et~al.}(2021)Weight, Gifford, Tretiak, and Kilina]{weight_interplay_2021}
B.~M. Weight, B.~J. Gifford, S.~Tretiak and S.~Kilina, \emph{The Journal of Physical Chemistry C}, 2021, \textbf{125}, 4785--4793\relax
\mciteBstWouldAddEndPuncttrue
\mciteSetBstMidEndSepPunct{\mcitedefaultmidpunct}
{\mcitedefaultendpunct}{\mcitedefaultseppunct}\relax
\EndOfBibitem
\bibitem[Setaro \emph{et~al.}(2017)Setaro, Adeli, Glaeske, Przyrembel, Bisswanger, Gordeev, Maschietto, Faghani, Paulus, Weinelt, Arenal, Haag, and Reich]{setaro_preserving_2017}
A.~Setaro, M.~Adeli, M.~Glaeske, D.~Przyrembel, T.~Bisswanger, G.~Gordeev, F.~Maschietto, A.~Faghani, B.~Paulus, M.~Weinelt, R.~Arenal, R.~Haag and S.~Reich, \emph{Nature Communications}, 2017, \textbf{8}, 14281\relax
\mciteBstWouldAddEndPuncttrue
\mciteSetBstMidEndSepPunct{\mcitedefaultmidpunct}
{\mcitedefaultendpunct}{\mcitedefaultseppunct}\relax
\EndOfBibitem
\bibitem[Hayashi \emph{et~al.}(2022)Hayashi, Niidome, Shiga, Yu, Nakagawa, Janas, Fujigaya, and Shiraki]{hayashi_azide_2022}
K.~Hayashi, Y.~Niidome, T.~Shiga, B.~Yu, Y.~Nakagawa, D.~Janas, T.~Fujigaya and T.~Shiraki, \emph{Chemical Communications}, 2022, \textbf{58}, 11422--11425\relax
\mciteBstWouldAddEndPuncttrue
\mciteSetBstMidEndSepPunct{\mcitedefaultmidpunct}
{\mcitedefaultendpunct}{\mcitedefaultseppunct}\relax
\EndOfBibitem
\bibitem[Sander \emph{et~al.}(2024)Sander, Metternich, Dippner, Kruss, and Borchardt]{sander_controlled_2024}
M.~Sander, J.~Metternich, P.~Dippner, S.~Kruss and L.~Borchardt, \emph{Controlled {Introduction} of sp3 quantum defects in {Fluorescent} {Carbon} {Nanotubes} by {Mechanochemistry}}, 2024, \url{https://chemrxiv.org/engage/chemrxiv/article-details/671f2dee83f22e42146a3ad6}\relax
\mciteBstWouldAddEndPuncttrue
\mciteSetBstMidEndSepPunct{\mcitedefaultmidpunct}
{\mcitedefaultendpunct}{\mcitedefaultseppunct}\relax
\EndOfBibitem
\bibitem[Karousis \emph{et~al.}(2010)Karousis, Tagmatarchis, and Tasis]{karousis_current_2010}
N.~Karousis, N.~Tagmatarchis and D.~Tasis, \emph{Chemical Reviews}, 2010, \textbf{110}, 5366--5397\relax
\mciteBstWouldAddEndPuncttrue
\mciteSetBstMidEndSepPunct{\mcitedefaultmidpunct}
{\mcitedefaultendpunct}{\mcitedefaultseppunct}\relax
\EndOfBibitem
\bibitem[Gifford \emph{et~al.}(2018)Gifford, Kilina, Htoon, Doorn, and Tretiak]{gifford_exciton_2018}
B.~J. Gifford, S.~Kilina, H.~Htoon, S.~K. Doorn and S.~Tretiak, \emph{The Journal of Physical Chemistry C}, 2018, \textbf{122}, 1828--1838\relax
\mciteBstWouldAddEndPuncttrue
\mciteSetBstMidEndSepPunct{\mcitedefaultmidpunct}
{\mcitedefaultendpunct}{\mcitedefaultseppunct}\relax
\EndOfBibitem
\bibitem[Settele \emph{et~al.}(2021)Settele, Berger, Lindenthal, Zhao, El~Yumin, Zorn, Asyuda, Zharnikov, Högele, and Zaumseil]{settele_synthetic_2021}
S.~Settele, F.~J. Berger, S.~Lindenthal, S.~Zhao, A.~A. El~Yumin, N.~F. Zorn, A.~Asyuda, M.~Zharnikov, A.~Högele and J.~Zaumseil, \emph{Nature Communications}, 2021, \textbf{12}, 2119\relax
\mciteBstWouldAddEndPuncttrue
\mciteSetBstMidEndSepPunct{\mcitedefaultmidpunct}
{\mcitedefaultendpunct}{\mcitedefaultseppunct}\relax
\EndOfBibitem
\bibitem[Liu \emph{et~al.}(2023)Liu, Shan, Wei, Wen, Jiang, Hu, Fang, Tang, and Li]{liu_novel_2023}
Z.~Liu, C.~Shan, G.~Wei, J.~Wen, L.~Jiang, G.~Hu, Z.~Fang, T.~Tang and M.~Li, \emph{Molecules}, 2023, \textbf{28}, 3637\relax
\mciteBstWouldAddEndPuncttrue
\mciteSetBstMidEndSepPunct{\mcitedefaultmidpunct}
{\mcitedefaultendpunct}{\mcitedefaultseppunct}\relax
\EndOfBibitem
\bibitem[Luo \emph{et~al.}(2023)Luo, Sun, Chi, Chai, Sun, and Wu]{luo_comparative_2023}
K.~Luo, W.~Sun, Y.~Chi, S.~Chai, C.~Sun and W.~Wu, \emph{Journal of Molecular Structure}, 2023, \textbf{1294}, 136525\relax
\mciteBstWouldAddEndPuncttrue
\mciteSetBstMidEndSepPunct{\mcitedefaultmidpunct}
{\mcitedefaultendpunct}{\mcitedefaultseppunct}\relax
\EndOfBibitem
\bibitem[Ma \emph{et~al.}(2022)Ma, Ning, and Wei]{ma_s-doped_2022}
G.~Ma, G.~Ning and Q.~Wei, \emph{Carbon}, 2022, \textbf{195}, 328--340\relax
\mciteBstWouldAddEndPuncttrue
\mciteSetBstMidEndSepPunct{\mcitedefaultmidpunct}
{\mcitedefaultendpunct}{\mcitedefaultseppunct}\relax
\EndOfBibitem
\bibitem[Jin \emph{et~al.}(2025)Jin, Lee, Lim, Lee, Park, Jung, Kang, and Na]{jin_sulfur-doped_2025}
M.~Jin, S.~Lee, S.~b. Lim, M.~Lee, J.~Park, H.-D. Jung, M.-H. Kang and K.~Na, \emph{Small}, 2025, \textbf{21}, 2410765\relax
\mciteBstWouldAddEndPuncttrue
\mciteSetBstMidEndSepPunct{\mcitedefaultmidpunct}
{\mcitedefaultendpunct}{\mcitedefaultseppunct}\relax
\EndOfBibitem
\bibitem[Li \emph{et~al.}(2025)Li, Mihm, Chen, Hou, Wen, Sharifzadeh, and Ma]{li_near-infrared_2025}
X.~Li, T.~N. Mihm, J.-S. Chen, H.~Hou, J.~Wen, S.~Sharifzadeh and X.~Ma, \emph{The Journal of Physical Chemistry C}, 2025, \textbf{129}, 17590--17598\relax
\mciteBstWouldAddEndPuncttrue
\mciteSetBstMidEndSepPunct{\mcitedefaultmidpunct}
{\mcitedefaultendpunct}{\mcitedefaultseppunct}\relax
\EndOfBibitem
\bibitem[Chen \emph{et~al.}(2017)Chen, Yin, Li, Cen, Li, and Yin]{chen_curvature_2017}
Y.~Chen, S.~Yin, Y.~Li, W.~Cen, J.~Li and H.~Yin, \emph{Applied Surface Science}, 2017, \textbf{404}, 364--369\relax
\mciteBstWouldAddEndPuncttrue
\mciteSetBstMidEndSepPunct{\mcitedefaultmidpunct}
{\mcitedefaultendpunct}{\mcitedefaultseppunct}\relax
\EndOfBibitem
\bibitem[Yu and Yi(2007)]{yu_single-walled_2007}
S.~Yu and W.~Yi, \emph{IEEE Transactions on Nanotechnology}, 2007, \textbf{6}, 545--548\relax
\mciteBstWouldAddEndPuncttrue
\mciteSetBstMidEndSepPunct{\mcitedefaultmidpunct}
{\mcitedefaultendpunct}{\mcitedefaultseppunct}\relax
\EndOfBibitem
\bibitem[Mittal and Kumar(2014)]{mittal_carbon_2014}
M.~Mittal and A.~Kumar, \emph{Sensors and Actuators B: Chemical}, 2014, \textbf{203}, 349--362\relax
\mciteBstWouldAddEndPuncttrue
\mciteSetBstMidEndSepPunct{\mcitedefaultmidpunct}
{\mcitedefaultendpunct}{\mcitedefaultseppunct}\relax
\EndOfBibitem
\bibitem[Yao \emph{et~al.}(2011)Yao, Duong, Lim, Yang, Hwang, Yu, Lee, Güneş, and Lee]{yao_humidity-assisted_2011}
F.~Yao, D.~L. Duong, S.~C. Lim, S.~B. Yang, H.~R. Hwang, W.~J. Yu, I.~H. Lee, F.~Güneş and Y.~H. Lee, \emph{Journal of Materials Chemistry}, 2011, \textbf{21}, 4502--4508\relax
\mciteBstWouldAddEndPuncttrue
\mciteSetBstMidEndSepPunct{\mcitedefaultmidpunct}
{\mcitedefaultendpunct}{\mcitedefaultseppunct}\relax
\EndOfBibitem
\bibitem[Goldoni \emph{et~al.}(2003)Goldoni, Larciprete, Petaccia, and Lizzit]{goldoni_single-wall_2003}
A.~Goldoni, R.~Larciprete, L.~Petaccia and S.~Lizzit, \emph{Journal of the American Chemical Society}, 2003, \textbf{125}, 11329--11333\relax
\mciteBstWouldAddEndPuncttrue
\mciteSetBstMidEndSepPunct{\mcitedefaultmidpunct}
{\mcitedefaultendpunct}{\mcitedefaultseppunct}\relax
\EndOfBibitem
\bibitem[Peymani \emph{et~al.}(2016)Peymani, Izadyar, and Nakhaeipour]{peymani_functionalization_2016}
S.~Peymani, M.~Izadyar and A.~Nakhaeipour, \emph{Physical Chemistry Research}, 2016, \textbf{4}, 553--565\relax
\mciteBstWouldAddEndPuncttrue
\mciteSetBstMidEndSepPunct{\mcitedefaultmidpunct}
{\mcitedefaultendpunct}{\mcitedefaultseppunct}\relax
\EndOfBibitem
\bibitem[Oftadeh \emph{et~al.}(2014)Oftadeh, Gholamian, and Abdallah]{oftadeh_sulfur_2014}
M.~Oftadeh, M.~Gholamian and H.~H. Abdallah, \emph{Physical Chemistry Research}, 2014, \textbf{2}, 30--40\relax
\mciteBstWouldAddEndPuncttrue
\mciteSetBstMidEndSepPunct{\mcitedefaultmidpunct}
{\mcitedefaultendpunct}{\mcitedefaultseppunct}\relax
\EndOfBibitem
\bibitem[Shen \emph{et~al.}(2014)Shen, Li, Liu, and Yin]{shen_dependence_2014}
W.~Shen, F.~Li, C.~Liu and L.-C. Yin, \emph{Chemical Physics Letters}, 2014, \textbf{608}, 1--5\relax
\mciteBstWouldAddEndPuncttrue
\mciteSetBstMidEndSepPunct{\mcitedefaultmidpunct}
{\mcitedefaultendpunct}{\mcitedefaultseppunct}\relax
\EndOfBibitem
\bibitem[Kresse and Furthmüller(1996)]{kresse_efficient_1996}
G.~Kresse and J.~Furthmüller, \emph{Physical Review B}, 1996, \textbf{54}, 11169--11186\relax
\mciteBstWouldAddEndPuncttrue
\mciteSetBstMidEndSepPunct{\mcitedefaultmidpunct}
{\mcitedefaultendpunct}{\mcitedefaultseppunct}\relax
\EndOfBibitem
\bibitem[Kresse and Furthmüller(1996)]{kresse_efficiency_1996}
G.~Kresse and J.~Furthmüller, \emph{Computational Materials Science}, 1996, \textbf{6}, 15--50\relax
\mciteBstWouldAddEndPuncttrue
\mciteSetBstMidEndSepPunct{\mcitedefaultmidpunct}
{\mcitedefaultendpunct}{\mcitedefaultseppunct}\relax
\EndOfBibitem
\bibitem[Kresse and Hafner(1993)]{kresse_ab_1993}
G.~Kresse and J.~Hafner, \emph{Physical Review B}, 1993, \textbf{47}, 558--561\relax
\mciteBstWouldAddEndPuncttrue
\mciteSetBstMidEndSepPunct{\mcitedefaultmidpunct}
{\mcitedefaultendpunct}{\mcitedefaultseppunct}\relax
\EndOfBibitem
\bibitem[Kresse and Hafner(1994)]{kresse_ab_1994}
G.~Kresse and J.~Hafner, \emph{Physical Review B}, 1994, \textbf{49}, 14251--14269\relax
\mciteBstWouldAddEndPuncttrue
\mciteSetBstMidEndSepPunct{\mcitedefaultmidpunct}
{\mcitedefaultendpunct}{\mcitedefaultseppunct}\relax
\EndOfBibitem
\bibitem[Kresse and Hafner(1994)]{kresse_norm-conserving_1994}
G.~Kresse and J.~Hafner, \emph{Journal of Physics: Condensed Matter}, 1994, \textbf{6}, 8245--8257\relax
\mciteBstWouldAddEndPuncttrue
\mciteSetBstMidEndSepPunct{\mcitedefaultmidpunct}
{\mcitedefaultendpunct}{\mcitedefaultseppunct}\relax
\EndOfBibitem
\bibitem[Blöchl(1994)]{blochl_projector_1994}
P.~E. Blöchl, \emph{Physical Review B}, 1994, \textbf{50}, 17953--17979\relax
\mciteBstWouldAddEndPuncttrue
\mciteSetBstMidEndSepPunct{\mcitedefaultmidpunct}
{\mcitedefaultendpunct}{\mcitedefaultseppunct}\relax
\EndOfBibitem
\bibitem[Kresse and Joubert(1999)]{kresse_ultrasoft_1999}
G.~Kresse and D.~Joubert, \emph{Physical Review B}, 1999, \textbf{59}, 1758--1775\relax
\mciteBstWouldAddEndPuncttrue
\mciteSetBstMidEndSepPunct{\mcitedefaultmidpunct}
{\mcitedefaultendpunct}{\mcitedefaultseppunct}\relax
\EndOfBibitem
\bibitem[Perdew \emph{et~al.}(1996)Perdew, Burke, and Ernzerhof]{perdew_generalized_1996}
J.~P. Perdew, K.~Burke and M.~Ernzerhof, \emph{Physical Review Letters}, 1996, \textbf{77}, 3865--3868\relax
\mciteBstWouldAddEndPuncttrue
\mciteSetBstMidEndSepPunct{\mcitedefaultmidpunct}
{\mcitedefaultendpunct}{\mcitedefaultseppunct}\relax
\EndOfBibitem
\bibitem[Grimme \emph{et~al.}(2010)Grimme, Antony, Ehrlich, and Krieg]{grimme_consistent_2010}
S.~Grimme, J.~Antony, S.~Ehrlich and H.~Krieg, \emph{The Journal of Chemical Physics}, 2010, \textbf{132}, 154104\relax
\mciteBstWouldAddEndPuncttrue
\mciteSetBstMidEndSepPunct{\mcitedefaultmidpunct}
{\mcitedefaultendpunct}{\mcitedefaultseppunct}\relax
\EndOfBibitem
\bibitem[He \emph{et~al.}(2014)He, Liu, Hautier, Oliveira, Marques, Vila, Rehr, Rignanese, and Zhou]{he_accuracy_2014}
L.~He, F.~Liu, G.~Hautier, M.~J.~T. Oliveira, M.~A.~L. Marques, F.~D. Vila, J.~J. Rehr, G.-M. Rignanese and A.~Zhou, \emph{Physical Review B}, 2014, \textbf{89}, 064305\relax
\mciteBstWouldAddEndPuncttrue
\mciteSetBstMidEndSepPunct{\mcitedefaultmidpunct}
{\mcitedefaultendpunct}{\mcitedefaultseppunct}\relax
\EndOfBibitem
\bibitem[Haas \emph{et~al.}(2009)Haas, Tran, and Blaha]{haas_calculation_2009}
P.~Haas, F.~Tran and P.~Blaha, \emph{Physical Review B}, 2009, \textbf{79}, 085104\relax
\mciteBstWouldAddEndPuncttrue
\mciteSetBstMidEndSepPunct{\mcitedefaultmidpunct}
{\mcitedefaultendpunct}{\mcitedefaultseppunct}\relax
\EndOfBibitem
\bibitem[Krukau \emph{et~al.}(2006)Krukau, Vydrov, Izmaylov, and Scuseria]{krukau_influence_2006}
A.~V. Krukau, O.~A. Vydrov, A.~F. Izmaylov and G.~E. Scuseria, \emph{The Journal of Chemical Physics}, 2006, \textbf{125}, 224106\relax
\mciteBstWouldAddEndPuncttrue
\mciteSetBstMidEndSepPunct{\mcitedefaultmidpunct}
{\mcitedefaultendpunct}{\mcitedefaultseppunct}\relax
\EndOfBibitem
\bibitem[He \emph{et~al.}(2017)He, Gifford, Hartmann, Ihly, Ma, Kilina, Luo, Shayan, Strauf, Blackburn, Tretiak, Doorn, and Htoon]{he_low-temperature_2017}
X.~He, B.~J. Gifford, N.~F. Hartmann, R.~Ihly, X.~Ma, S.~V. Kilina, Y.~Luo, K.~Shayan, S.~Strauf, J.~L. Blackburn, S.~Tretiak, S.~K. Doorn and H.~Htoon, \emph{ACS Nano}, 2017, \textbf{11}, 10785--10796\relax
\mciteBstWouldAddEndPuncttrue
\mciteSetBstMidEndSepPunct{\mcitedefaultmidpunct}
{\mcitedefaultendpunct}{\mcitedefaultseppunct}\relax
\EndOfBibitem
\bibitem[Hilmer \emph{et~al.}(2012)Hilmer, McNicholas, Lin, Zhang, Wang, Mendenhall, Song, Heller, Barone, Blankschtein, and Strano]{hilmer_role_2012}
A.~J. Hilmer, T.~P. McNicholas, S.~Lin, J.~Zhang, Q.~H. Wang, J.~D. Mendenhall, C.~Song, D.~A. Heller, P.~W. Barone, D.~Blankschtein and M.~S. Strano, \emph{Langmuir}, 2012, \textbf{28}, 1309--1321\relax
\mciteBstWouldAddEndPuncttrue
\mciteSetBstMidEndSepPunct{\mcitedefaultmidpunct}
{\mcitedefaultendpunct}{\mcitedefaultseppunct}\relax
\EndOfBibitem
\bibitem[Goldoni \emph{et~al.}(2004)Goldoni, Petaccia, Gregoratti, Kaulich, Barinov, Lizzit, Laurita, Sangaletti, and Larciprete]{goldoni_spectroscopic_2004}
A.~Goldoni, L.~Petaccia, L.~Gregoratti, B.~Kaulich, A.~Barinov, S.~Lizzit, A.~Laurita, L.~Sangaletti and R.~Larciprete, \emph{Carbon}, 2004, \textbf{42}, 2099--2112\relax
\mciteBstWouldAddEndPuncttrue
\mciteSetBstMidEndSepPunct{\mcitedefaultmidpunct}
{\mcitedefaultendpunct}{\mcitedefaultseppunct}\relax
\EndOfBibitem
\bibitem[Yu and Wang(2023)]{yu_excitons_2023}
G.~Yu and L.~Wang, \emph{Physica B: Condensed Matter}, 2023, \textbf{667}, 415143\relax
\mciteBstWouldAddEndPuncttrue
\mciteSetBstMidEndSepPunct{\mcitedefaultmidpunct}
{\mcitedefaultendpunct}{\mcitedefaultseppunct}\relax
\EndOfBibitem
\bibitem[Biktagirov \emph{et~al.}(2025)Biktagirov, Gerstmann, and Schmidt]{biktagirov_topological_2025}
T.~Biktagirov, U.~Gerstmann and W.~G. Schmidt, \emph{Nanoscale}, 2025, \textbf{17}, 6884--6891\relax
\mciteBstWouldAddEndPuncttrue
\mciteSetBstMidEndSepPunct{\mcitedefaultmidpunct}
{\mcitedefaultendpunct}{\mcitedefaultseppunct}\relax
\EndOfBibitem
\bibitem[Humeres and Moreira(2012)]{humeres_kinetics_2012}
E.~Humeres and R.~d. F. P.~M. Moreira, \emph{Journal of Physical Organic Chemistry}, 2012, \textbf{25}, 1012--1026\relax
\mciteBstWouldAddEndPuncttrue
\mciteSetBstMidEndSepPunct{\mcitedefaultmidpunct}
{\mcitedefaultendpunct}{\mcitedefaultseppunct}\relax
\EndOfBibitem
\bibitem[Humeres \emph{et~al.}(2017)Humeres, Debacher, Moreira, Santaballa, and Canle]{humeres_reactive_2017}
E.~Humeres, N.~A. Debacher, R.~d. F. P.~M. Moreira, J.~A. Santaballa and M.~Canle, \emph{The Journal of Physical Chemistry C}, 2017, \textbf{121}, 14649--14657\relax
\mciteBstWouldAddEndPuncttrue
\mciteSetBstMidEndSepPunct{\mcitedefaultmidpunct}
{\mcitedefaultendpunct}{\mcitedefaultseppunct}\relax
\EndOfBibitem
\bibitem[Shokuhi~Rad \emph{et~al.}(2016)Shokuhi~Rad, Esfahanian, Maleki, and Gharati]{shokuhi_rad_application_2016}
A.~Shokuhi~Rad, M.~Esfahanian, S.~Maleki and G.~Gharati, \emph{Journal of Sulfur Chemistry}, 2016, \textbf{37}, 176--188\relax
\mciteBstWouldAddEndPuncttrue
\mciteSetBstMidEndSepPunct{\mcitedefaultmidpunct}
{\mcitedefaultendpunct}{\mcitedefaultseppunct}\relax
\EndOfBibitem
\bibitem[noa()]{noauthor_covalent_nodate}
\emph{Covalent {Bond} {Energies}}, \url{https://gchem.cm.utexas.edu/data/section2.php?target=bond-energies-table4.php}\relax
\mciteBstWouldAddEndPuncttrue
\mciteSetBstMidEndSepPunct{\mcitedefaultmidpunct}
{\mcitedefaultendpunct}{\mcitedefaultseppunct}\relax
\EndOfBibitem
\bibitem[noa(2013)]{noauthor_fundamentals_2013}
\emph{Fundamentals of {Chemical} {Bonding}}, 2013, \url{https://chem.libretexts.org/Bookshelves/Physical_and_Theoretical_Chemistry_Textbook_Maps/Supplemental_Modules_(Physical_and_Theoretical_Chemistry)/Chemical_Bonding/Fundamentals_of_Chemical_Bonding}\relax
\mciteBstWouldAddEndPuncttrue
\mciteSetBstMidEndSepPunct{\mcitedefaultmidpunct}
{\mcitedefaultendpunct}{\mcitedefaultseppunct}\relax
\EndOfBibitem
\bibitem[Haynes(2016)]{haynes_crc_2016}
W.~Haynes, \emph{{CRC} {Handbook} of {Chemistry} and {Physics}}, CRC Press, 2016\relax
\mciteBstWouldAddEndPuncttrue
\mciteSetBstMidEndSepPunct{\mcitedefaultmidpunct}
{\mcitedefaultendpunct}{\mcitedefaultseppunct}\relax
\EndOfBibitem
\bibitem[Ruzsinszky \emph{et~al.}(2006)Ruzsinszky, Perdew, Csonka, Vydrov, and Scuseria]{ruzsinszky_spurious_2006}
A.~Ruzsinszky, J.~P. Perdew, G.~I. Csonka, O.~A. Vydrov and G.~E. Scuseria, \emph{The Journal of Chemical Physics}, 2006, \textbf{125}, 194112\relax
\mciteBstWouldAddEndPuncttrue
\mciteSetBstMidEndSepPunct{\mcitedefaultmidpunct}
{\mcitedefaultendpunct}{\mcitedefaultseppunct}\relax
\EndOfBibitem
\end{mcitethebibliography}
 \end{document}